\documentclass[aps,prl,twocolumn,amsfonts,nofootinbib,notitlepage,preprintnumbers]{revtex4-1}
\usepackage{hyperref}
\usepackage{mathrsfs}
\hypersetup{
colorlinks=true,
linkcolor=red,
citecolor=blue,
urlcolor=blue}
\usepackage{epsf,epsfig}
\usepackage{amscd}
\usepackage{amsmath,amssymb,bm}
\usepackage{subfig}
\usepackage{graphicx}
\usepackage[countmax]{subfloat}
\input xy
\xyoption{all} 

\newcommand{\bea}{\begin{eqnarray}}
\newcommand{\eea}{\end{eqnarray}}
\newcommand{\nn}{\nonumber}
\renewcommand\d{\partial}

\begin{document}

\author{Srimoyee Sen}
\affiliation{Department of Physics, The University of Arizona, Tucson, Arizona 85721, USA} 

\author{Naoki Yamamoto}
\affiliation{Department of Physics, Keio University, Yokohama 223-8522, Japan}

\title{Chiral Shock Waves}

\begin{abstract}
We study the shock waves in relativistic chiral matter. 
We argue that the conventional Rankine-Hugoinot relations are modified due to the presence of 
chiral transport phenomena. We show that the entropy discontinuity in a weak shock wave is 
quadratic in the pressure discontinuity when the effect of chiral transport becomes sufficiently large. 
We also show that rarefaction shock waves, which do not exist in usual nonchiral fluids, can appear 
in chiral matter. The direction of shock wave propagation is found to be completely determined by 
the direction of the vorticity and the chirality of fermions. These features are exemplified by shock 
propagation in dense neutrino matter in the hydrodynamic regime.
\end{abstract}

\maketitle
\emph{Introduction.}---%
Recently, relativistic chiral matter has attracted great interest both theoretically and 
experimentally. Chiral matter is considered to be realized in a wide range of systems from 
the electroweak plasmas in the early Universe \cite{Joyce:1997uy,Boyarsky:2011uy},
quark-gluon plasmas in heavy ion collisions \cite{Kharzeev:2007jp,Fukushima:2008xe}, 
Weyl (semi)metals \cite{Nielsen:1983rb,Vishwanath,BurkovBalents,Xu-chern}, and 
electron plasmas in neutron stars \cite{Charbonneau:2009ax,Ohnishi:2014uea,Kaminski:2014jda} 
to neutrino media in core-collapse supernova explosions \cite{Yamamoto:2015gzz,Yamamoto:2016xtu}.
A remarkable property of chiral matter is the presence of unusual transport phenomena related 
to quantum anomalies in field theory \cite{Adler,BellJackiw}, called the chiral magnetic effect in 
a magnetic field \cite{Vilenkin:1980fu,Nielsen:1983rb,Alekseev:1998ds,Fukushima:2008xe} 
and chiral vortical effect (CVE) in a vorticity \cite{Vilenkin:1979ui,Kharzeev:2007tn,Son:2009tf,Landsteiner:2011cp}.
These chiral transport phenomena lead to new types of collective modes, such as the chiral 
magnetic wave \cite{Newman:2005hd,Kharzeev:2010gd}, chiral vortical wave \cite{Jiang:2015cva}, 
chiral Alfv\'en wave \cite{Yamamoto:2015ria}, chiral heat wave \cite{Chernodub:2015gxa}, 
and the chiral plasma instability \cite{Joyce:1997uy,Boyarsky:2011uy,Akamatsu:2013pjd};
see also Refs.~\cite{Abbasi:2015saa, Kalaydzhyan:2016dyr} for recent related works.

In this Letter, we study the shock propagation in relativistic chiral matter. We first argue that the 
so-called Rankine-Hugoniot relations, or the jump conditions at the shock front \cite{Landau}, 
must be modified by the presence of chiral transport phenomena. This in turn leads to 
modifications of the basic properties of weak shock waves in chiral matter. 
Our main findings are summarized as follows.
\begin{itemize}
\item{The dependence of entropy discontinuity at the shock front on the corresponding pressure 
discontinuity is quadratic, $\Delta S \propto \left(\Delta p\right)^2$ [see Eq.~(\ref{dS_chiral})], 
when the effect of chiral transport becomes sufficiently large. This should be contrasted with the 
behavior $\Delta S \propto (\Delta p)^3$ in nonchiral matter \cite{Landau}. }
\item{Rarefaction shock waves can appear in chiral matter for a sufficiently large vorticity
$\omega \gg \omega_{\rm c}$, where $\omega_{\rm c}$ is defined in Eq.~(\ref{cond}). 
This should be contrasted with the fact that rarefaction shock waves are usually prohibited in 
nonchiral matter \cite{Landau, note}.}
\item{For a given chirality of fermions, the direction of shock wave propagation is completely 
determined by the direction of the vorticity.}
\end{itemize}
We exemplify these features by studying shock waves in dense charge neutral chiral matter in the 
hydrodynamic regime. Such a situation is realized, e.g., by neutrino media at the core of supernovae 
\cite{Yamamoto:2015gzz}. These qualitatively new aspects of shock waves in chiral matter may 
have possible relevance, e.g., to the dynamics of supernovae. Although our argument does in part 
depend on this particular case of chiral matter, we expect that the qualitative features of our result 
are more generic. We discuss the possible applications of our arguments and results to other 
systems in the conclusion. In this Letter, we use the metric $g^{\mu \nu} = {\rm diag}(1,-1,-1,-1)$.

\emph{Chiral hydrodynamics.}---%
Our starting point is the relativistic chiral hydrodynamics \cite{Son:2009tf}. With keeping a 
specific application to the neutrino media in supernovae \cite{Yamamoto:2015gzz} in mind, 
we consider chiral hydrodynamics for charge neutral chiral matter. Our argument can be 
extended to charged chiral matter in external electromagnetic fields in a straightforward 
manner. For simplicity, we will here ignore the dissipative effects.

The equations of relativistic chiral hydrodynamics for a single (right- or left-handed) chiral fermion 
are given by energy-momentum conservation and particle number conservation as
\bea
\partial_{\mu} T^{\mu \nu} &=& 0, \\
\partial_{\mu} j^{\mu} &=& 0.
\eea
Here the energy-momentum tensor $T^{\mu \nu}$ and particle number current $j^{\mu}$ are 
given in the Landau-Lifshitz frame by \cite{Son:2009tf}
\bea
\label{Tmunu}
T^{\mu \nu} &= & h u^{\mu} u^{\nu} - p g^{\mu \nu},
\\
\label{jmu}
j^{\mu} & = & nu^{\mu} + \xi \omega^{\mu},
\eea
where $h = \epsilon + p$ is the enthalpy density, $n$ is the particle number density, 
$u^{\mu} = \gamma (1, {\bm v})$ is the fluid velocity with $\gamma = (1-{\bm v}^2)^{-1/2}$, and 
$\omega^{\mu}=\frac{1}{2}\epsilon^{\mu\nu\lambda\rho}u_{\nu}\partial_{\lambda}u_{\rho}$ is the fluid vorticity. 

What is different in the neutral chiral hydrodynamics from the conventional nonchiral hydrodynamics 
\cite{Landau} is the presence of the CVE proportional to $\omega^{\mu}$ in Eq.~(\ref{jmu}) 
\cite{Vilenkin:1979ui,Kharzeev:2007tn,Son:2009tf,Landsteiner:2011cp}. 
The transport coefficient $\xi$ takes the form of \cite{Son:2009tf,Landsteiner:2011cp,Neiman:2010zi}
\bea
\label{xi}
\xi= C \mu^2 \left(1-\frac{2}{3}\frac{n\mu}{h} \right) + D T^2 \left(1-\frac{2n\mu}{h} \right)\,,
\eea
where $\mu$ is the chemical potential and $T$ is the temperature. The coefficients $C$ and $D$ 
are related to those of the chiral anomaly and mixed gauge-gravitational anomaly as 
\cite{Son:2009tf,Landsteiner:2011cp,Golkar:2012kb,Jensen:2012kj}
\bea
\label{CD}
C = \pm \frac{1}{4\pi^2}\,, \qquad D= \pm \frac{1}{12}\,,
\eea
for right- and left-handed chiral fermions, respectively.

In the following, we will focus on the regime $\mu \gg T$ for demonstration. This is relevant to, 
e.g.,  the dense neutrino matter at the core of supernovae \cite{Yamamoto:2015gzz}. Note 
however that our argument itself is not limited to this regime and is applicable to other regimes 
as well. For a relativistic gas of noninteracting fermions (which is a reasonable assumption for 
a neutrino gas), the expressions of $n$, $p$, $\epsilon$, and the entropy $S$ are given by
\begin{align}
\label{n}
n &= \frac{\mu^3}{6\pi^2}+ \frac{\mu T^2}{6}\,, \\
\label{p}
p &= \frac{\epsilon}{3} = \frac{\mu^4}{24 \pi^2} + \frac{\mu^2 T^2}{12}\,, \\ 
\label{S}
S &= \frac{\pi^2 T}{\mu}\,, 
\end{align}
to the leading corrections in ${T}/{\mu} \ll 1$. From Eqs.~(\ref{n}) and (\ref{p}), the transport 
coefficient $\xi$ in Eq.~(\ref{xi}) in the regime $\mu \gg T$ reduces to
\bea
\label{xi_approx}
\xi \approx \frac{1}{3} C \mu^2 + \left(\frac{2\pi^2}{3}C - D \right)T^2\,.
\eea

\emph{Shock waves in relativistic nonchiral matter.}---%
We first revisit the properties of shock waves in relativistic nonchiral matter \cite{Landau} before 
we analyze shock waves in chiral matter. Consider a relativistic gas of particles moving along the 
$x$ axis towards the positive $x$ direction such that there is a surface of discontinuity 
perpendicular to the direction of propagation of the gas. This surface of discontinuity divides the 
three-dimensional space into two regions, side $1$ and side $2$. The sides are defined in such 
a way that the gas moves from side $1$ to side $2$. 

Imposing continuity in particle number flux, energy flux, and momentum flux, 
$j^x_1 = j^x_2$, $T^{xx}_1 = T^{xx}_2$, and $T^{0x}_1 = T^{0x}_2$, 
we have the following three equations relating the two sides \cite{Landau}:
\begin{gather}
\label{jx_non}
\frac{v_1\gamma_1}{V_1}=\frac{v_2\gamma_2}{V_2}\,, \\
\label{Txx_non}
h_1v_1^2\gamma_1^2+p_1=h_2v_2^2\gamma_2^2+p_2, \\
\label{T0x_non}
h_1v_1\gamma_1^2=h_2v_2\gamma_2^2,
\end{gather}
where the subscripts $1$ and $2$ stand for sides $1$ and $2$ and $V$ is the volume per 
particle, $V\equiv {1}/{n}$. Equations~(\ref{jx_non})--(\ref{T0x_non}) constitute the 
Rankine-Hugoniot relations for shock propagation in nonchiral matter.

By solving Eqs.~(\ref{Txx_non}) and (\ref{T0x_non}) in terms of $v_1$ and $v_2$, we have
\cite{Landau}
\begin{align}
\label{v1}
v_1&=\sqrt{\frac{(p_2-p_1)(\epsilon_2+p_1)}{(\epsilon_2-\epsilon_1)(\epsilon_1+p_2)}}\,, \\
\label{v2}
v_2&=\sqrt{\frac{(p_2-p_1)(\epsilon_1+p_2)}{(\epsilon_2-\epsilon_1)(\epsilon_2+p_1)}}\,.
\end{align}
Substituting these expressions into Eq.~(\ref{jx_non}), we obtain the pressure-volume 
relation \cite{Landau}:
\bea
h_1^2V_1^2-h_2^2V_2^2 + (p_2-p_1)(h_1V_1^2+h_2V_2^2) = 0.
\label{adiab}
\eea 
For given $p_1$ and $V_1$, it provides the relation between $p_2$ and $V_2$ under the 
equation of state $p = p(\epsilon)$.

Let us consider the case of weak shock waves. For weak shock waves, we have 
$\epsilon_2\rightarrow\epsilon_1$ and $p_2\rightarrow p_1$, etc. If we expand the expression 
for the speed on side $1$ given in Eq.~(\ref{v1}) in ${\Delta \epsilon}/{h_1} \ll 1$ and 
${\Delta p}/{h_1} \ll 1$ with $\Delta \epsilon \equiv \epsilon_2-\epsilon_1$ and 
$\Delta p \equiv p_2 - p_1$, we find that
\bea
(v_1)^2&=&\lim_{2\rightarrow1} \frac{\Delta p}{\Delta \epsilon } \frac{\epsilon_2+p_1}{\epsilon_1+p_2}\nn \\
&=&\frac{dp}{d\epsilon}\bigg|_1\lim_{2\rightarrow1}\left(1+\frac{\Delta \epsilon}{h_1}-\frac{\Delta p}{h_1}+\cdots\right)\nn\\
&=&\frac{dp}{d\epsilon}\bigg|_1\left[1+\frac{\Delta \epsilon}{h_1}\left(1-\frac{dp}{d\epsilon}\bigg|_1\right)+\cdots\right]\nn\\
&=& (c_{{\rm s}1})^2\left[1+\frac{\Delta \epsilon}{h_1}\left(1- (c_{{\rm s}1})^2 \right)+\cdots \right]\,,
\eea
where $c_{{\rm s}1}$ is the speed of sound on side $1$, $(c_{{\rm s}1})^2 = (dp/d\epsilon)_1$. 
Similarly we find that
\bea
(v_2)^2=(c_{{\rm s}2})^2\left[1-\frac{\Delta \epsilon}{h_1}\left(1-(c_{{\rm s}2})^2\right)+\cdots\right]\,,
\eea
where $c_{{\rm s}2}$ is the speed of sound on side $2$, $(c_{{\rm s}2})^2 = (dp/d\epsilon)_2$. 
In order to simplify our discussion below, we assume the equations of states on the two sides 
to be the same, where $c_{{\rm s}1}=c_{{\rm s}2}$.

As can be seen from Eqs.~(\ref{v1}) and (\ref{v2}), $(v_1)^2 > (v_2)^2$ when 
$\epsilon_2>\epsilon_1$, and $(v_1)^2 < (v_2)^2$ when $\epsilon_2<\epsilon_1$. 
The former is known as a compression shock wave and the latter is known as a rarefaction 
shock wave. Remember also that, for compression shock waves, we have $p_2>p_1$, and 
for rarefaction shock waves, we have $p_1>p_2$. This is ensured by the fact that
\bea
c_{\rm s}^2=\lim_{2\rightarrow1}\frac{p_2-p_1}{\epsilon_2-\epsilon_1} > 0\,. 
\eea
Although the equations of relativistic hydrodynamics allow for the existence of both compression 
shock waves and rarefaction shock waves, only compression shock waves are consistent with the 
second law of thermodynamics, and rarefaction shock waves are not. 

In order to see it, we expand the adiabatic of Eq.~(\ref{adiab}) using
\bea
\Delta H &=& T \Delta S +V_1 \Delta p+\frac{1}{2}\frac{\d V}{\d p}\bigg|_1 \! \! (\Delta p)^2
+\frac{1}{6}\frac{\d^2 V}{\d p^2}\bigg|_1 \! \! (\Delta p)^3 + \cdots\,, 
\nn \\
\label{exp1}
\\
\Delta V &=& \frac{\d V}{\d p}\bigg|_{1}\Delta p + \frac{1}{2}\frac{\d^2 V}{\d p^2}\bigg|_1 \! \! (\Delta p)^2
+\frac{1}{6}\frac{\d^3 V}{\d p^3}\bigg|_1 \! \! (\Delta p)^3 \nn \\
& & +\frac{\d V}{\d S}\bigg|_{1}\Delta S + \cdots\,, 
\label{exp2}
\eea
the former of which can be derived by using the thermodynamic relation $V = \left.{\d H}/{\d p}\right|_S$.
Here, $H =h V$ is the enthalpy, $S = s V$ is the entropy, and we defined $\Delta H = H_2-H_1$, 
$\Delta S = S_2-S_1$, and $\Delta V = V_2-V_1$. 
After substituting Eqs.~(\ref{exp1}) and (\ref{exp2}) into Eq.~(\ref{adiab}), one finds \cite{Landau}
\bea
\label{dS}
\Delta S = \frac{1}{12 H_1 T}\frac{\d^2 (HV)}{\d p^2}\bigg|_1(\Delta p)^3 + O\! \left((\Delta p)^4 \right)\,. 
\eea

It can be seen by plotting the adiabatic for any reasonable equation of state that $\d^2(HV)/\d p^2 >0$; 
for ultrarelativistic gases at low $T$, see the explicit calculation of the coefficient of the term 
$(\Delta p)^3$ in Eq.~(\ref{dS_general}) below. The second law of thermodynamics requires that 
$S_2>S_1$, which then requires that $p_2>p_1$. This corresponds to compression shock waves 
explained earlier. Hence, we see that, in nonchiral relativistic fluids, the second law of 
thermodynamics allows for the existence of compression shock waves alone and that there are 
no rarefaction shock waves. This is the so-called Zempl\'{e}n theorem (see also \cite{note}).

\emph{Chiral shock waves.}---%
Let us now consider shock waves in chiral matter in the presence of a vorticity $\omega^{\mu}$.
The important point here is that the vorticity induces the CVE in chiral matter, leading to the 
modifications of the Rankine-Hugoniot relations. As a result, the basic properties of shock waves are 
qualitatively modified compared with those in nonchiral matter above. 

We again consider a situation where a shock wave is moving along the $x$ axis towards the 
positive $x$ direction. For simplicity, we consider the case with a nonzero constant vorticity in 
the $x$ direction,
\bea
\omega_{x}=\frac{1}{2}u_{t}(\partial_z u_y-\partial_y u_z) \equiv \omega\,,
\label{omega1}
\eea
with $\omega_y = \omega_z = 0$,
but the extension of our analysis to more general cases should be straightforward. Note that we 
are in a system where we cannot go to a frame where $v_y = v_z = 0$ at all points, unlike the 
case of shock waves in nonchiral matter discussed earlier. 

In what follows, we consider the regime $|\omega| \rho \ll 1$, where $\rho = \sqrt{y^2 + z^2}$ is 
the distance from the axis of the vorticity. Assuming that $\partial_y v_x = \partial_z v_x =0$, we 
can solve Eq.~(\ref{omega1}) for $v_{\perp} \equiv \sqrt{v_y^2+v_z^2}$ to find
\bea
\label{vperp}
v_{\perp} = \omega \rho (1-v_x^2) + O\! \left((\omega \rho)^2 \right).
\eea

Now, let us look at how the jump conditions at the shock front are modified in the presence of the 
vorticity. Continuity of particle flux and energy-momentum flux across a surface of discontinuity, 
$j^x_1 = j^x_2$, $T^{xx}_1 = T^{xx}_2$, $T^{0x}_1 = T^{0x}_2$, $T^{yx}_1 = T^{yx}_2$,
and $T^{zx}_1 = T^{zx}_2$, reads
\bea
\label{jx}
\frac{v_1^x\gamma_1}{V_1} + \xi_1\omega_1 &=& \frac{v_2^x\gamma_2}{V_2}+\xi_2\omega_2 \equiv j \,, \\
\label{Txx}
h_1 (v^x_1)^2\gamma_1^2 + p_1 &=& h_2 (v^x_2)^2\gamma_2^2 + p_2, \\
\label{Tx0}
h_1 v^x_1\gamma_1^2 &=& h_2 v^x_2\gamma_2^2, \\
\label{Txy}
h_1 v_1^{x}v_1^{y}\gamma_1^2 &=& h_2 v_2^{x}v_2^{y} \gamma_2^2, \\
\label{Txz}
h_1 v_1^{x}v_1^{z}\gamma_1^2 &=& h_2 v_2^{x}v_2^{z} \gamma_2^2.
\label{hydeqv}
\eea
The modification, compared with Eqs.~(\ref{jx_non})--(\ref{T0x_non}) in nonchiral matter, 
is the presence of the CVE in Eq.~(\ref{jx}).

Notice first that Eqs.~(\ref{Txy}) and (\ref{Txz}) require that $v_1^{\perp}=v_2^{\perp}$. 
From Eq.~(\ref{vperp}), we then must have $\omega_1 [1-(v_1^x)^2]=\omega_2 [1-(v_2^x)^2]$. 
Below we will be interested in weak shock waves, for which $v_2^x \rightarrow v_1^x$. In this limit
\bea 
\label{domega}
\Delta \omega \equiv \omega_2-\omega_1 =\omega_1\frac{(v_2^x)^2-(v_1^x)^2}{1-c_{{\rm s}1}^2}\,,
\eea
up to terms that are suppressed by $|(v_1^x)^2-(v_2^x)^2| \ll 1$.
To analyze the weak shock wave, only Eqs.~(\ref{jx})--(\ref{Tx0}) are thus relevant, which 
constitute the modified Rankine-Hugoniot relations for shock propagation in chiral matter. 

These equations may also be viewed as the interface conditions between the two sides of chiral 
matter in a {\it global} rotation ${\bm \Omega} = \Omega \hat {\bm x}$ under the replacement 
$\omega \rightarrow \Omega$.

From Eqs.~(\ref{jx}) and (\ref{Txx}), we have
\bea
(h_1 V_1^2 - h_2 V_2^2)j^2 - 2  (h_1 V_1^2 \omega_1\xi_1 -  h_2 V_2^2 \omega_2\xi_2) j \nn \\
+ p_1-p_2 +  (h_1 \omega_1^2\xi_1^2 V_1^2 - h_2 \omega_2^2\xi_2^2 V_2^2) = 0\,.
\eea
This can be solved for $j$ as
\bea
\label{j}
j=\frac{(h_1V_1^2\xi_1\omega_1-h_2V_2^2\xi_2\omega_2) \pm \sqrt{\cal D}}{h_1V_1^2-h_2V_2^2}
\eea
with 
\bea
{\cal D} \equiv (p_2-p_1)(h_1V_1^2-h_2V_2^2)+h_1 h_2 V_1^2 V_2^2 (\xi_1\omega_1-\xi_2\omega_2)^2.
\nn \\
\eea
In order for $j$ to be real, we must have ${\cal D} \geq 0$. 

Note that we must have $ |\omega_{1,2}| \ll \mu_{1,2}$ for hydrodynamics to make sense, where 
$\mu_{1,2}$ is the chemical potential of side $1$ or $2$. Since we are interested in the regime 
$\frac{4\pi}{3} \rho^3 \gg V$, the limit $|\omega_{1,2}| \rho \ll 1$ is within the applicability of 
hydrodynamics. In this limit, Eqs.~(\ref{Txx}) and (\ref{Tx0}) can be solved to obtain the 
expressions for $v_1^x$ and $v_2^x$, which invariably turn out to be identical to the expressions 
in Eqs.~(\ref{v1}) and (\ref{v2}) up to corrections suppressed by powers of $\rho \omega_{1,2}$. Hence, 
we can use the leading-order expression for the velocities in the expansion of $\rho \omega_{1,2}$ in 
Eq.~(\ref{jx}) to obtain the relation on the two sides as
\bea
\label{adiab_chiral}
\frac{v_1}{V_1\sqrt{1-v_1^2}} - \frac{v_2}{V_2\sqrt{1-v_2^2}} =-(\omega_1\xi_1 - \omega_2\xi_2) \,.
\eea
Here, $v_{1,2}$ are given in Eqs.~(\ref{v1}) and (\ref{v2}).
Recalling that $\mu_{1,2}$ and $T_{1,2}$ can be written in terms of $p_{1,2}$ and $V_{1,2}$
from Eqs.~(\ref{n}) and (\ref{p}), both sides of Eq.~(\ref{adiab_chiral}) can be expressed in terms 
of $p_{1,2}$ and $V_{1,2}$ alone, for a given equation of state $p=p(\epsilon)$. This provides 
the pressure-volume relation of the two sides for chiral matter in a constant vorticity. 

Let us now consider how the relation (\ref{dS}) is modified for weak shock waves in chiral matter. 
The modified relation will be obtained by expanding Eq.~(\ref{adiab_chiral}) in terms of  
$\Delta S = S_2-S_1$ and $\Delta p = p_2-p_1$ across the surface of discontinuity. The 
expansion of the left-hand side of Eq.~(\ref{adiab_chiral}), but ignoring the right-hand side, 
would lead to the result (\ref{dS}) for nonchiral matter. Here, we need to expand the right-hand 
side in $\Delta S$ and $\Delta p$ as well.

For this purpose, we first solve $p=p(\mu, T)$ and $S=S(\mu,T)$ in Eqs.~(\ref{p}) and (\ref{S}) 
for $\mu$ and $T$ in terms of $p$ and $S$, by treating $S \ll 1$ as a perturbation for $\mu \gg T$. 
The results to the leading correction in $S \ll 1$ are given by
\begin{align}
\mu(p, S) &= \left(1-\frac{S^2}{2\pi^2} \right)(24 \pi^2 p)^{1/4} \,, \\
T(p, S) &= \frac{S}{\pi^2}(24 \pi^2 p)^{1/4} \,.
\end{align}
From these equations, $\Delta \mu \equiv \mu_2 -\mu_1$ and $\Delta T \equiv T_2 - T_1$ can 
be expanded in $\Delta p$ and $\Delta S$ as
\begin{align}
\label{dmu}
\Delta \mu & 
= \frac{6\pi^2}{\mu^3}\Delta p -T\Delta S \,, \\
\label{dT}
\Delta T & 
= \frac{6\pi^2 T}{\mu^4}\Delta p + \frac{\mu}{\pi^2}\Delta S \,.
\end{align}

We then expand $\Delta \xi \equiv \xi_2 -\xi_1$ in $\Delta S$ and $\Delta p$. When $\mu \gg T$,
using Eqs.~(\ref{CD}), (\ref{xi_approx}), (\ref{dmu}), and (\ref{dT}), we have
\bea
\label{dxi}
\Delta \xi = \lambda \left[\frac{1}{\mu^2} \Delta p - \frac{6\pi^2}{\mu^6} (\Delta p)^2 + \cdots \right]\,,
\eea
where 
\bea
\lambda \equiv 4 \pi^2 C = \pm 1,
\eea
for right- and left-handed chiral matter, respectively. In Eq.~(\ref{dxi}), $``\cdots"$ stands for 
higher order terms in $\Delta p$ and terms of quadratic and higher order in $\Delta S$. Note here 
that the coefficient of $\Delta S$ in Eq.~(\ref{dxi}) is proportional to ${2C}/{3} - {2D}/{\pi^2}$ and 
vanishes identically regardless of the chirality.

Collecting the leading terms in the expansion of Eqs.~(\ref{adiab_chiral}) in $\Delta S$ and 
$\Delta p$ using Eqs.~(\ref{exp1}), (\ref{exp2}), (\ref{domega}), and (\ref{dxi}), we arrive at
\bea
\label{dS_general}
\Delta S=\frac{216 \pi^6}{\mu_1^{11}T_1}(\Delta p)^3 - \omega_1 \lambda \frac{36\sqrt{2}\pi^4}{\mu_1^8 T_1} (\Delta p)^2 + \cdots\,,
\eea
where $``\cdots"$ includes terms that are higher order in $\Delta p$. Note that the first term on the 
right-hand side of Eq.~(\ref{dS_general}) corresponds to the result (\ref{dS}) for nonchiral matter. 
The correction in the second term due to the vorticity becomes negligible and reduces 
to the result (\ref{dS}) when $|\omega_1| \ll \omega_{\rm c}$ for a given $\Delta p$, where
\bea
\label{cond}
\omega_{\rm c} = 3 \sqrt{2} \pi^2 \frac{\Delta p}{\mu_1^{3}}\,.
\eea

Let us now consider the case with $|\omega_1| \gg \omega_{\rm c}$. However, $\omega_1$ still 
needs to be small enough such that $|\omega_1| \rho \ll 1$ remains valid. In this case, 
Eq.~(\ref{dS_general}) is approximately
\bea
\label{dS_chiral}
\Delta  S \approx -\omega_1\lambda \frac{36\sqrt{2}\pi^4}{\mu_1^8 T_1} (\Delta p)^2 \,.
\eea
This is our main result.

Recall that the second law of thermodynamics requires that $\Delta S > 0$. From Eq.~(\ref{dS_chiral}), 
this allows for shock waves only when $\omega_1 \lambda < 0$. According to our conventions here, 
a positive vorticity is assumed to be pointing opposite to the direction of the shock wave propagation. 
Hence, for right-handed fermions (where $\lambda = 1$), we must have $\omega_1<0$ for shock 
wave propagation and vice versa. In particular, for neutrino matter (where $\lambda=-1$), we must have 
$\omega_1>0$ to satisfy $\Delta S>0$. The other remarkable feature of our result is that the sign of 
$\Delta S$ is independent of the sign of $\Delta p$. Hence, for $\omega_1 \lambda < 0$, both compression 
and rarefaction shock waves are realizable in contrast with nonchiral shock waves, which can only be of 
the compression type.

\emph{Discussion.}---%
In this Letter, we have explored the shock propagation in relativistic chiral matter. We have 
seen that the conventional Rankine-Hugoniot relations are modified due to the presence of 
chiral transport phenomena. In particular, we have shown that rarefaction shock waves can 
appear in chiral matter. Also we have shown that the existence of a shock wave itself is dependent 
on the chirality of the fermions involved and the direction of shock wave propagation. In this sense, 
the shock wave found in this Letter is {\it chiral}, similar to the other chiral waves 
\cite{Newman:2005hd,Kharzeev:2010gd,Jiang:2015cva,Yamamoto:2015ria,Chernodub:2015gxa}. 
It would be interesting to study possible phenomenological consequences of the chiral shock 
wave, e.g., in the dynamics of supernovae. In fact, the mean free path of neutrinos
$l_{\rm mfp} \lesssim 1 \ {\rm m}$ at the core of supernovae (with matter density 
$\rho \gtrsim 10^{13} \ {\rm g}/{\rm cm}^3$) is much smaller than the typical size of the core, 
$l_{\rm core} \sim 100 \ {\rm km}$. Hence, chiral hydrodynamics is applicable to neutrino matter 
at least in such a system \cite{Yamamoto:2015gzz}.

Although we limit ourselves to dense and cold charge neutral chiral matter in this Letter, our 
argument should be applicable to hot and/or dense charged chiral matter in an external magnetic 
field as well. In this case, the chiral magnetic effect modifies the Rankine-Hugoniot relations in 
a way qualitatively similar to what we found in this Letter. Such a new type of shock wave may 
be relevant to the electroweak plasmas in the early Universe.

This work was supported by JSPS KAKENHI Grants No.~16K17703 and the
MEXT-Supported Program for the Strategic Research Foundation at Private 
Universities, ``Topological Science'' (Grant No.~S1511006), and by the U.S. 
Department of Energy through Grant No.~DE-FG02-04ER41338.

{\it Note added.}---Recently, M.~N.~Chernodub informed the authors that he also obtained 
results \cite{Chernodub} that have some overlap with ours.


\begin{thebibliography}{99}
\bibitem{note}
  Precisely speaking, rarefaction shock waves can appear in nonchiral matter when 
  $\d^2 (HV)/\d p^2 |_S < 0$ (or $\d^2 V/\d p^2 |_S < 0$ in the nonrelativistic limit). 
  However, this condition is usually not met except in very particular situations, e.g., near 
  a gas-liquid critical point \cite{rarefaction}. On the other hand, the presence of chiral 
  rarefaction shock waves found in this Letter does not depend on the sign of 
  $\d^2 (HV)/\d p^2 |_S$, providing a novel realization of rarefaction shock waves.
\end{thebibliography}
\end{document}